\documentclass[superscriptaddress,amsmath,amssymb,onecolumn,amsfonts]{revtex4-2}
\usepackage{graphicx}
\graphicspath{{Pictures/}}
\usepackage{hyperref}
\hypersetup{pdfnewwindow=true, colorlinks=true, linkcolor=blue, anchorcolor=blue, citecolor=red, filecolor=blue, menucolor=blue, urlcolor=magenta}
\usepackage{cleveref}
\usepackage{color}
\usepackage{float}
\usepackage{soul}
\usepackage{ulem}
\usepackage{braket}
\usepackage{enumitem}
\usepackage{marginnote}

\newcommand{\sn}{{\rm sn}}
\newcommand{\fsd}{{\rm sd}}
\newcommand{\cd}{{\rm cd}}
\newcommand{\nd}{{\rm nd}}
\newcommand{\dn}{{\rm dn}}
\newcommand{\cn}{{\rm cn}}
\newcommand{\sech}{{\rm sech}}

\begin{document}

\title{New static solutions of symmetric $\phi^4$ equation}

\author{Avinash Khare}
\email{avinashkhare45@gmail.com }
\affiliation{Physics Department, Savitribai Phule Pune University, Pune 411007, India}
\author{Saikat Banerjee}
\email{saikatb@lanl.gov}
\affiliation{Theoretical Division, T-4, Los Alamos National Laboratory, Los Alamos, New Mexico 87545, USA}
\author{Avadh Saxena}
\email{avadh@lanl.gov}
\affiliation{Theoretical Division and Center for Nonlinear Studies, Los Alamos National Laboratory, Los Alamos, New Mexico 87545, USA}
\date{\today}

\begin{abstract}
In this paper, we provide new exact solutions of nonlinear Klein-Gordon ($\phi^4$) equation in $1+1$-dimension. For simplicity, we focus on the static equation and ignore the time-dependence. The symmetric $\phi^4$ equation has played an important role in several areas of physics. We obtain several novel non-singular solutions of the symmetric $\phi^4$ model in terms of the Jacobi elliptic functions and compare them with the well-known solutions. Finally, we categorize these solutions in terms of the potential parameters. 
\end{abstract}

\maketitle

\section{Introduction}
 
The symmetric $\phi^4$ model is one of the most celebrated models in physics. This model has played an important role in several different areas of physics. It represents one of the simplest models for second-order phase transitions~\cite{kle}. It has also played an important role in field theory~\cite{2}. The kink solution of this model~\cite{raj} is perhaps the most celebrated solution in the nonlinear literature. It is well known that there is a direct connection between the solutions of SU(2) Yang-Mills theory and symmetric $\phi^4$ model in one space dimension~\cite{kle,thooft,cof,wil,jac,act}. It is thus clearly of interest to discover new static solutions of the symmetric $\phi^4$  model in one space dimension characterized by the field equation
\begin{equation}\label{eq.1}
\phi_{xx}(x) = a \phi(x) + b\phi^3(x).
\end{equation}
It may be noted that being a Lorentz invariant model, knowing any static solution, we automatically know the corresponding time-dependent solution. So far as we are aware, apart from the celebrated kink solution $\tanh(x)$, the other well-known solutions of this model are pulse solution $\sech(x)$ and spatially periodic solutions~\cite{aubry,actor1} given by periodic kink solution $\sn(x,m)$ and periodic pulse solutions $\dn(x,m), \cn(x,m)$ where $0 \le m \le 1$ is the modulus of the Jacobi elliptic functions $\sn(x,m)$, $\cn(x,m)$, and $\dn(x,m)$~\cite{abr}. The nice property of these periodic solutions is that in the $m = 1$ limit, they go over to the well-known (hyperbolic) kink and pulse solutions. Subsequently, few more periodic solutions like $\nd(x,m), \fsd(x,m), \cd(x,m)$~\cite{abr,ks18} and superposed solution $\dn(x,m)\pm \cn(x,m)$ as well as complex parity-time-invariant ($\cal{PT}$) periodic solutions with $\cal{PT}$-eigenvalue +$1$ like $\dn(x,m) \pm i\sn(x,m), \cn(x,m) \pm i\sn(x,m)$ and $\cd(x,m) \pm i\fsd(x,m)$, and solutions with $\cal{PT}$-eigenvalue $-1$ like $\sn(x,m) \pm i\dn(x,m)$, $\sn(x,m) \pm i\cn(x,m)$ and $\fsd(x,m) \pm i\cd(x,m)$ were discovered by us~\cite{ks13,ks14,ks16,ks18}. Recently, we have also discovered four periodic solutions of this equation of the form  $\tfrac{A\sn(x,m)}{1+B\cn^2(x,m)}$, $\tfrac{A\cn(x,m)\dn(x,m)}{1+B\cn^2(x,m)}$, $\tfrac{A\dn(x,m)}{1+B\cn^2(x,m)}$, and $\tfrac{A\cn(x,m)\sn(x,m)}{1+B\cn^2(x,m)}$ which can be re-expressed as a superposition of two periodic kink solutions or two periodic $\dn$-type pulse solutions \cite{ks22}. We emphasize here that only non-singular solutions of Eq.~\eqref{eq.1} are considered in this paper.

The purpose of this work is to present several new periodic kink and pulse solutions, including complex $\cal{PT}$-invariant as well as complex but not $\cal{PT}$-invariant solutions of the static symmetric $\phi^4$ Eq.~\eqref{eq.1}. We make a detailed comparison of the new solutions with the well-known solutions mentioned above and show that, unlike the well-known solutions, many of the new solutions span a much wider space in the $a,b$ phase-space, where $a,b$ are the parameters of Eq.~\eqref{eq.1}.

\section{New Solutions \label{sec:sec_I}}

To begin, we will discuss the new periodic pulse solutions, then the periodic kink solutions followed by complex solutions, both $\cal{PT}$-invariant and $\cal{PT}$-noninvariant ones. \\

{\bf Solution I}

\begin{figure}[t]
\centering
\includegraphics[width=0.45\linewidth]{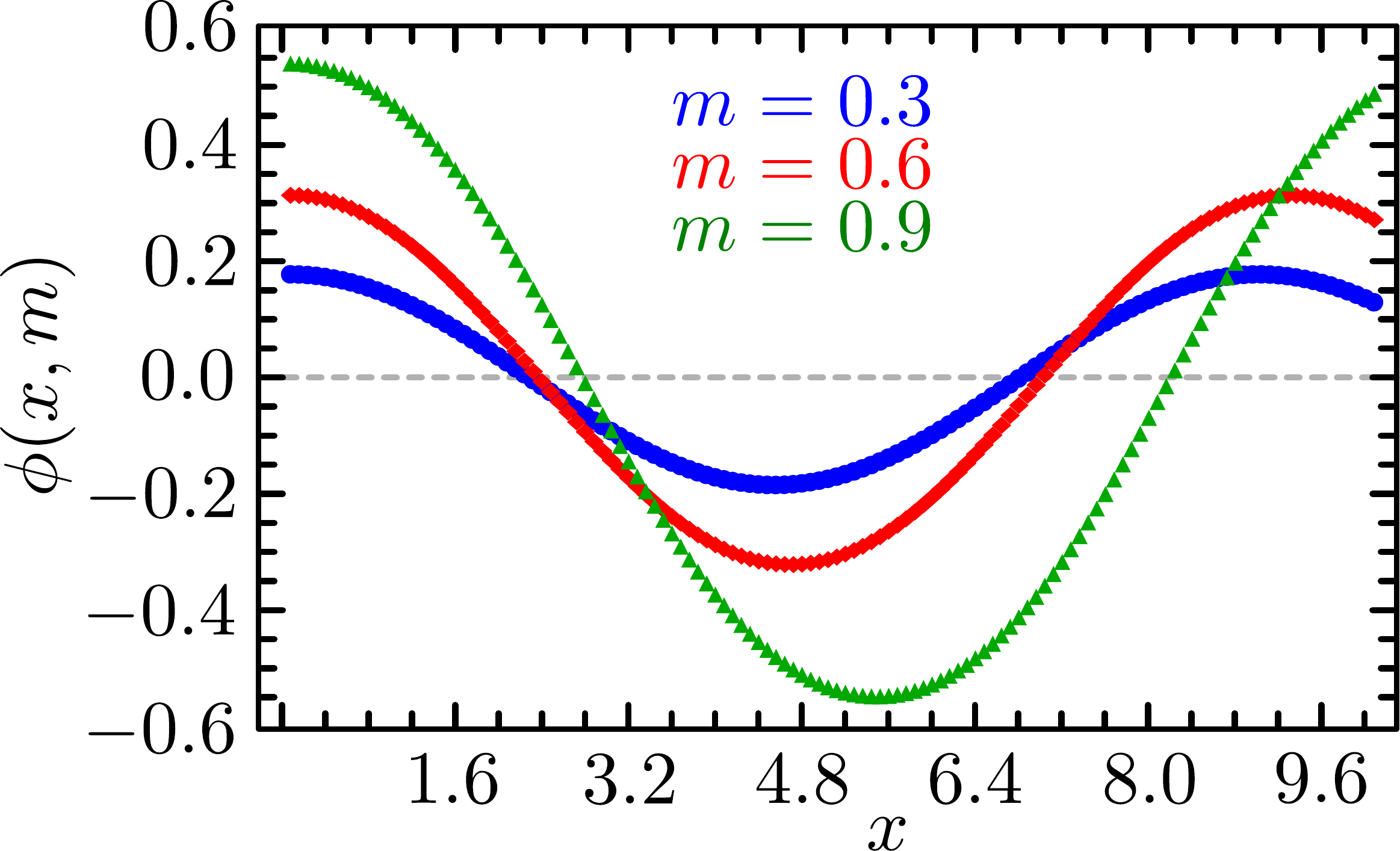} 
\caption{Variations of nonlinear Klein-Gordon fields $\phi(x,m)$ as a function of the position following Eq.~\eqref{eq.2}, and Eq.~\eqref{eq.3}. Note that the parameters are chosen as $a = - 0.5$, and $b = 0.8$, whereas the modulus variation is illustrated in different colors and symbols.}\label{fig:Fig1}
\end{figure}

It is easy to check that 
\begin{equation}\label{eq.2}
\phi(x) = \frac{A\sqrt{m}\cn(\beta x,m)}{D+\dn(\beta x,m)}, 
\qquad 
A,D > 0,
\end{equation}
is an exact periodic pulse solution of the symmetric $\phi^4$ Eq.~\eqref{eq.1} provided
\begin{equation}\label{eq.3}
D^2 = 1-m > 0,
\qquad
2 b A^2 = m \beta^2,
\qquad
a = -(2-m)\frac{\beta^2}{2}.
\end{equation}
Note that this solution exists only if $b > 0, a < 0$ and $0 < m < 1$. In terms of the parameters $a,b$ of the $\phi^4$ model, the solution can be re-expressed as
\begin{equation}\label{eq.4}
\phi(x) = \frac{\sqrt{\frac{|a|m}{(2-m)b}}\, \cn[\sqrt{\frac{2|a|}{(2-m)}} x,m]}{\sqrt{1-m}+\dn[\sqrt{\frac{2|a|}{(2-m)}} x,m]}.
\end{equation}
In contrast, the corresponding well-known periodic pulse solution given by 
\begin{equation}\label{eq.5}
\phi = A\cn(\beta x,m),
\qquad
A > 0\,
\end{equation}
exists provided 
\begin{equation}\label{eq.6}
a = (2m-1)\beta^2,
\qquad
b A^2 = -2\beta^2.
\end{equation}
Thus, while the new pulse solution I exists only if $b > 0, a < 0$, the well-known solution Eq.~\eqref{eq.5} exists only if $b < 0$ while $a$ could be positive or negative. Thus the two solutions span different parameter spaces in the $(a,b)$ plane. Note that in the limit $m = 1$, the solution Eq.~\eqref{eq.2} goes over to the constant solution $\phi = A$. Thus there is no hyperbolic pulse solution with the above ansatz. The structure of these solutions is illustrated in Fig.~\ref{fig:Fig1} for distinct choices of the 
modulus $m$, and parameters $a = - 0.5$, and $b = 0.8$.\\

{\bf Solution II}

It is easy to check that 
\begin{equation}\label{eq.7}
\phi = \frac{A\dn(\beta x,m)}{D+\sn(\beta x,m)},
\qquad
A > 0, D > 1,
\end{equation}
is an exact periodic pulse solution of $\phi^4$ Eq.~\eqref{eq.1} provided
\begin{equation}\label{eq.8}
0 < m < 1,
\qquad
m D^2 = 1,
\qquad
2 bm A^2 = -(1-m) \beta^2,
\qquad
a = (1+m)\frac{\beta^2}{2}.
\end{equation}
Note that this solution exists only if $b < 0, a > 0$. Like solution I, solution II can also be entirely re-expressed in terms of the parameters $a, b$ of the $\phi^4$ Eq.~\eqref{eq.1}. This feature is valid for all the solutions presented below.

On the other hand, the corresponding well-known periodic $\dn$-type pulse solution is
\begin{equation}\label{eq.9}
\phi = A \dn(\beta x, m),
\end{equation}
provided
\begin{equation}\label{eq.10}
b A^2 = -2\beta^2,
\qquad
a= (2-m)\beta^2.
\end{equation}
Hence, this solution also exists provided $b < 0, a > 0$. However, while the well-known solution Eq.~\eqref{eq.9} is invariant under $x \rightarrow -x$, the new Solution II is not so, and hence the two solutions have very different behavior in a period. Note that the new solution has period $4K(m)$ while the well-known solution Eq.~\eqref{eq.9} has period $2K(m)$ where $K(m)$ is the complete elliptic integral of the first kind~\cite{abr}. Note also that the solution II goes to zero in the limit $m = 1$. \\

{\bf Solution III}

\begin{figure}[t]
\centering
\includegraphics[width=0.45\linewidth]{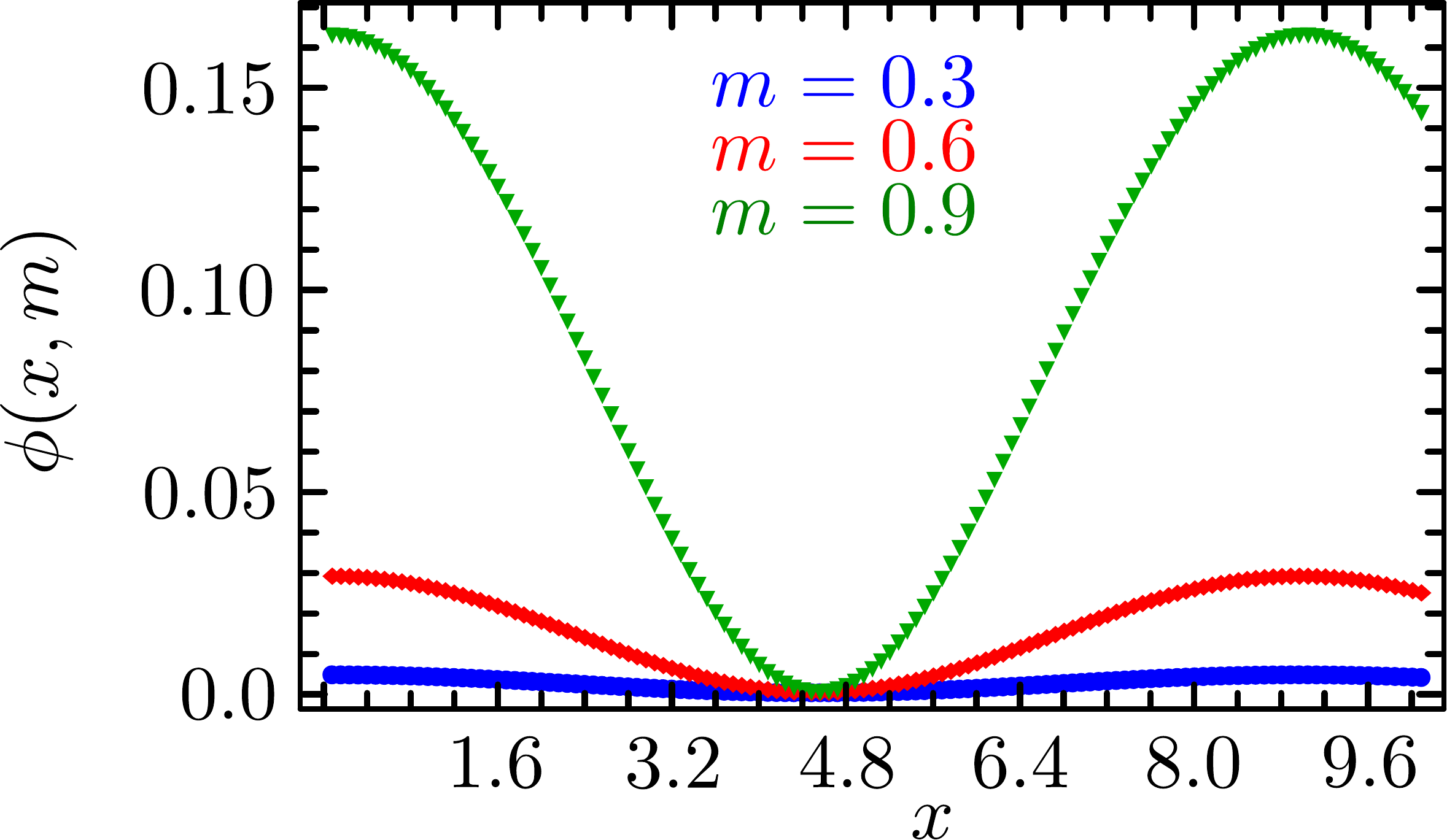} 
\caption{Variations of nonlinear Klein-Gordon fields $\phi(x,m)$ as a function of the position following Eq.~\eqref{eq.11}, Eq.~\eqref{eq.12.1}, and Eq.~\eqref{eq.12.2}. Note that the parameters are chosen as $a = - 0.5$ and $b = 0.8$.}\label{fig:Fig2}
\end{figure}

It is easy to check that 
\begin{equation}\label{eq.11}
\phi = \frac{A\dn(\beta x,m)}{D+\dn(\beta x,m)} - F,
\qquad
D,F,A > 0,
\end{equation}
is an exact periodic pulse solution of $\phi^4$ Eq.~\eqref{eq.1} provided
\begin{subequations}
\begin{align}
\label{eq.12.1}
D^2 & = \sqrt{1-m} > 0,
\qquad
b A^2 = 2(1-\sqrt{1-m})^2 \beta^2, \\
\label{eq.12.2}
F 	&	= A/2,
\qquad
a = -[(2-m)+6\sqrt{1-m}]\frac{\beta^2}{2}.
\end{align}
\end{subequations}
Note that in contrast to solution II and the well-known solution Eq.~\eqref{eq.9}, solution III exists only if $b > 0, a < 0$. Thus compared to the other two $\dn$-type pulse solutions, solution III covers an entirely different parameter space in the $(a,b)$ plane. Using the relations Eq.~\eqref{eq.12.1} and Eq.~\eqref{eq.12.2}, the solution Eq.~\eqref{eq.11} can be re-expressed as  
\begin{equation}\label{eq.13}
\phi = \frac{F[\dn(\beta x,m)-D]}{\dn(\beta x,m)+D},
\end{equation}
where
\begin{equation}\label{eq.14}
D = (1-m)^{1/4},
\qquad
F = \sqrt{\frac{|a|(1-\sqrt{1-m})^2}{b(2-m+6\sqrt{1-m})}}.
\end{equation}
The corresponding hyperbolic pulse solution however does not exist as in case 
$m = 1$, one has $A,F,D = 0$. The structure of these periodic pulse solutions is shown in Fig.~\ref{fig:Fig2} for various choices of the modulus $m$, and fixed parameters $a = - 0.5$, and $b  = 0.8$. \\

{\bf Solution IV}

It is easy to check that 
\begin{equation}\label{eq.15}
\phi = \frac{[A\dn(\beta x,m)+B\sqrt{m}\cn(\beta x,m)]}{D+\sn(\beta x,m)},
\qquad
A, B > 0, D > 1,
\end{equation}
is an exact superposed periodic solution of $\phi^4$ Eq.~\eqref{eq.1} provided
\begin{equation}\label{eq.16}
2 b A^2 = -(D^2-1)\beta^2,
\qquad
2 m b B^2 = -(m D^2-1)\beta^2,
\qquad
a = (1+m)\frac{\beta^2}{2}.
\end{equation}
Note that for this solution $a > 0, b < 0, m D^2 > 1$. The corresponding well known superposed pulse solution of the $\phi^4$ Eq.~\eqref{eq.1} is
\begin{equation}\label{eq.17}
\phi = [A\dn(\beta x,m)+B\sqrt{m}\cn(\beta x,m)],
\end{equation}
provided
\begin{equation}\label{eq.18}
B = \pm A,
\qquad
2 b A^2 = -\beta^2,
\qquad
a = (1+m)\frac{\beta^2}{2}.
\end{equation}
The two solutions IV and Eq.~\eqref{eq.17} are very different, while the solution Eq.~\eqref{eq.17} is invariant under $x \rightarrow -x$, this is not the case for the solution IV and hence their behavior is very different in one period. \\

{\bf Solution V}

It is well known that the $\phi^4$ Eq.~\eqref{eq.1} admits both $A\dn(x,m)$ and $B/\dn(x,m)$~\cite{ks18} as exact solutions, but we now show that remarkably, even their superposition is an exact solution of the $\phi^4$ Eq.~\eqref{eq.1}. In particular, it is easy to check that
\begin{equation}\label{eq.19}
\phi = A\dn(\beta x,m) + \frac{B\sqrt{1-m}}{\dn(\beta x,m)},
\end{equation}
is an exact solution of Eq.~\eqref{eq.1} provided
\begin{equation}\label{eq.20}
0 < m < 1,
\qquad
B = \pm A,
\qquad
b A^2 = -2\beta^2,
a = [(2-m)\pm 6\sqrt{1-m}]\beta^2.
\end{equation}
Note that the two $\pm$ signs are correlated. \\

{\bf Solution VI}

\begin{figure}[t]
\centering
\includegraphics[width=0.45\linewidth]{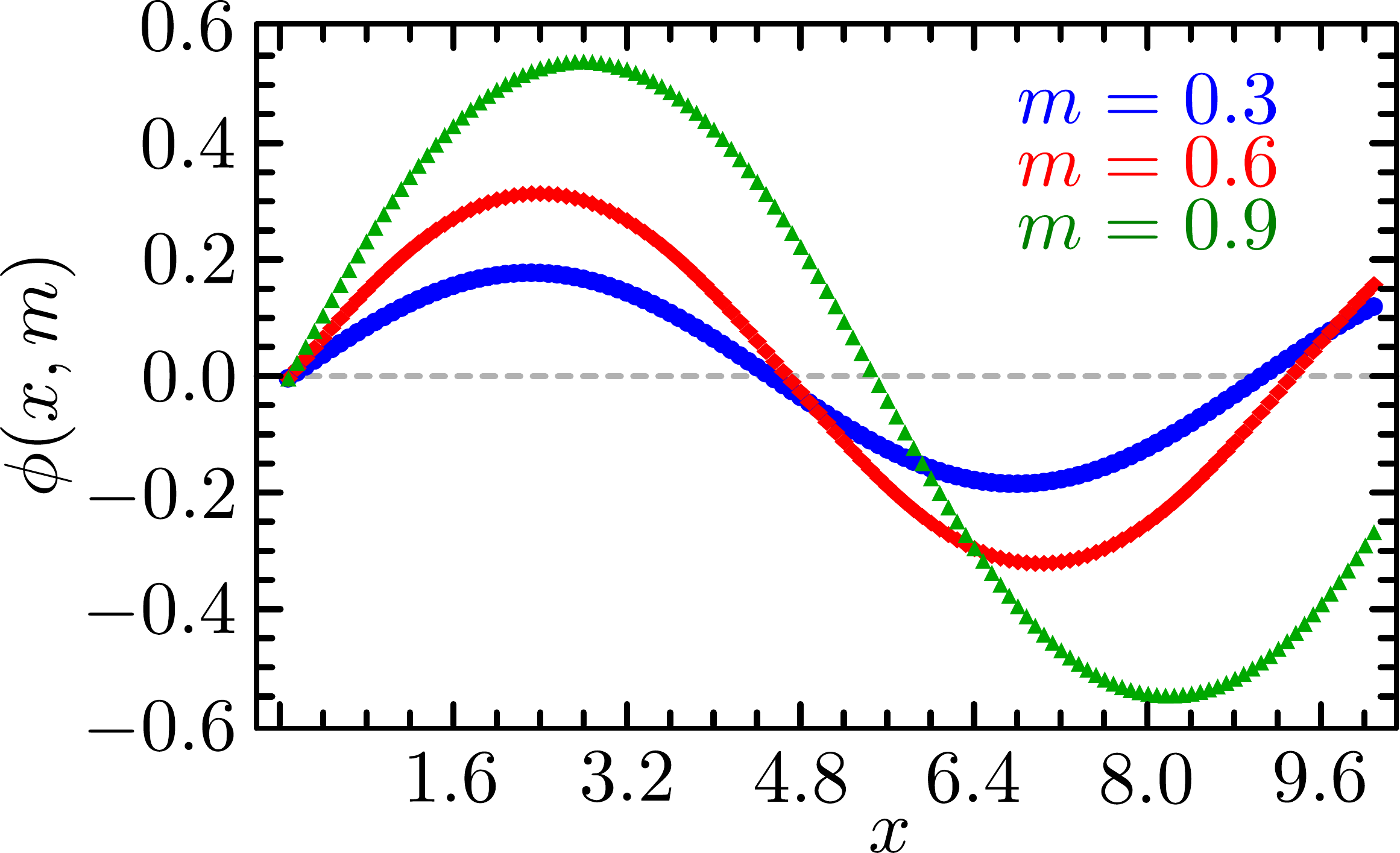} 
\caption{Variations of nonlinear Klein-Gordon fields $\phi(x,m)$ as a function of the position following Eq.~\eqref{eq.21}, and Eq.~\eqref{eq.22}. Note that the parameters are chosen as $a = - 0.5$ and $b = 0.8$. The variation with different modulus $m$ is illustrated in various colors and symbols.}\label{fig:Fig3}
\end{figure}

We now discuss three new periodic kink solutions. It is easy to check that 
\begin{equation}\label{eq.21}
\phi = \frac{A\sqrt{m}\sn(\beta x,m)}{D+\dn(\beta x,m)},
\qquad
A,D > 0,
\end{equation}
is an exact periodic kink solution of the $\phi^4$ Eq.~\eqref{eq.1} provided
\begin{equation}\label{eq.22}
D = 1,
\qquad
2 b A^2 = \beta^2,
\qquad
a = -(2-m)\frac{\beta^2}{2}.
\end{equation}
Thus this solution exists in case $a < 0, b > 0$. One should contrast this with the well-known periodic kink solution of $\phi^4$ Eq.~\eqref{eq.1}, i.e. 
\begin{equation}\label{eq.23}
\phi = A\sqrt{m}\sn(\beta x,m)
\end{equation}
provided
\begin{equation}\label{eq.24}
b A^2 = 2\beta^2,
\qquad
a = -(1+m)\beta^2.
\end{equation}
Note that, like solution VI, this solution holds if $a < 0, b > 0$. However, the two periodic kink solutions VI and Eq.~\eqref{eq.23} are very different. In the limit $m = 1$, the solution VI goes over to the hyperbolic kink solution
\begin{equation}\label{eq.25}
\phi = \frac{A\tanh(\beta x)}{D+\sech(\beta x)},
\qquad
A,D > 0,
\end{equation}
provided
\begin{equation}\label{eq.26}
D = 1,
\qquad
2 b A^2 = \beta^2,
\qquad
a = -\frac{\beta^2}{2}.
\end{equation}
Thus like the usual kink solution, this solution also exists when $a < 0, b > 0$. On the other hand, the conventional kink solution (which also follows from Eq.~\eqref{eq.23} and Eq.~\eqref{eq.24}) is
\begin{equation}\label{eq.27}
\phi = A \tanh(\beta x),
\end{equation}
provided
\begin{equation}\label{eq.28}
b A^2 = 2\beta^2,
\qquad
a = -2\beta^2.
\end{equation}
Note that this solution holds if $a < 0, b > 0$. The obvious question is, are the two kink solutions Eq.~\eqref{eq.25} and Eq.~\eqref{eq.27} distinct? The answer to this question is {\it no}. In particular, on using $\sinh(2x) = 2 \sinh(x) \cosh(x)$ and $1+\cosh(2x) = 2\cosh^2(x)$, it follows that the solutions Eq.~\eqref{eq.25}, and Eq.~\eqref{eq.27} are one and the same. The structure of these periodic kink solutions is shown in Fig.~\ref{fig:Fig3} for distinct choices of the modulus $m$ with the parameters $a = - 0.5$, and $b = 0.8$. \\

{\bf Solution VII}

It is easy to check that 
\begin{equation}\label{eq.29}
\phi = F-\frac{A\sn(\beta x,m)}{D+\sn(\beta x,m)},
\qquad
D > 1,
\end{equation}
is an exact periodic pulse solution of the $\phi^4$ Eq.~\eqref{eq.1} provided either $D^2 = 1/\sqrt{m}$ or $m =1$. Let us discuss both of the cases as follows: 

\begin{enumerate}[label=(\roman*)]
\item $D^2 = \tfrac{1}{\sqrt{m}}$: In this case, the periodic kink solution Eq.~\eqref{eq.29} exists provided
\begin{subequations}
\begin{align}
\label{eq.30.1}
A	&	=	2F,		&	b A^2	&	= -2(1-\sqrt{m})^2 \beta^2, \\
\label{eq.30.2}
0 	&	< m < 1, 	&	a 		&	= (6\sqrt{m}+1+m)\beta^2.
\end{align}
\end{subequations}
Thus, this periodic kink solution exists only if $b < 0, a > 0$.

\item $m = 1$: In this case, the solution Eq.~\eqref{eq.29} goes over to the hyperbolic kink solution
\begin{equation}\label{eq.31}
\phi = F-\frac{A\tanh(\beta x)}{D+\tanh(\beta x)},
\qquad
D > 1,
\end{equation}
provided
\begin{equation}\label{eq.32}
A = -(D^2-1)F,
\qquad
b A^2 D^2 = 2(D^2-1)^2 \beta^2,
\qquad
a = -2\beta^2.
\end{equation}
Thus in this case the solution Eq.~\eqref{eq.31} takes a simpler form
\begin{equation}\label{eq.33}
\phi = \frac{FD[1+D\tanh(\beta x)]}{D+\tanh(\beta x)}.
\end{equation}
Notice that as $x$ goes from $-\infty$ to +$\infty$, $\phi$ goes from $-FD$ to +$FD$ while at $x = 0$, $\phi = F$. Is this a new kink solution? The answer is {\it no}. It trivially follows from the usual kink solution $A\tanh(\beta x)$ by the translation $\beta x \rightarrow \beta x +a$ where $\tanh(a) = 1/D$. \\
\end{enumerate}

\begin{figure}[t]
\centering
\includegraphics[width=0.45\linewidth]{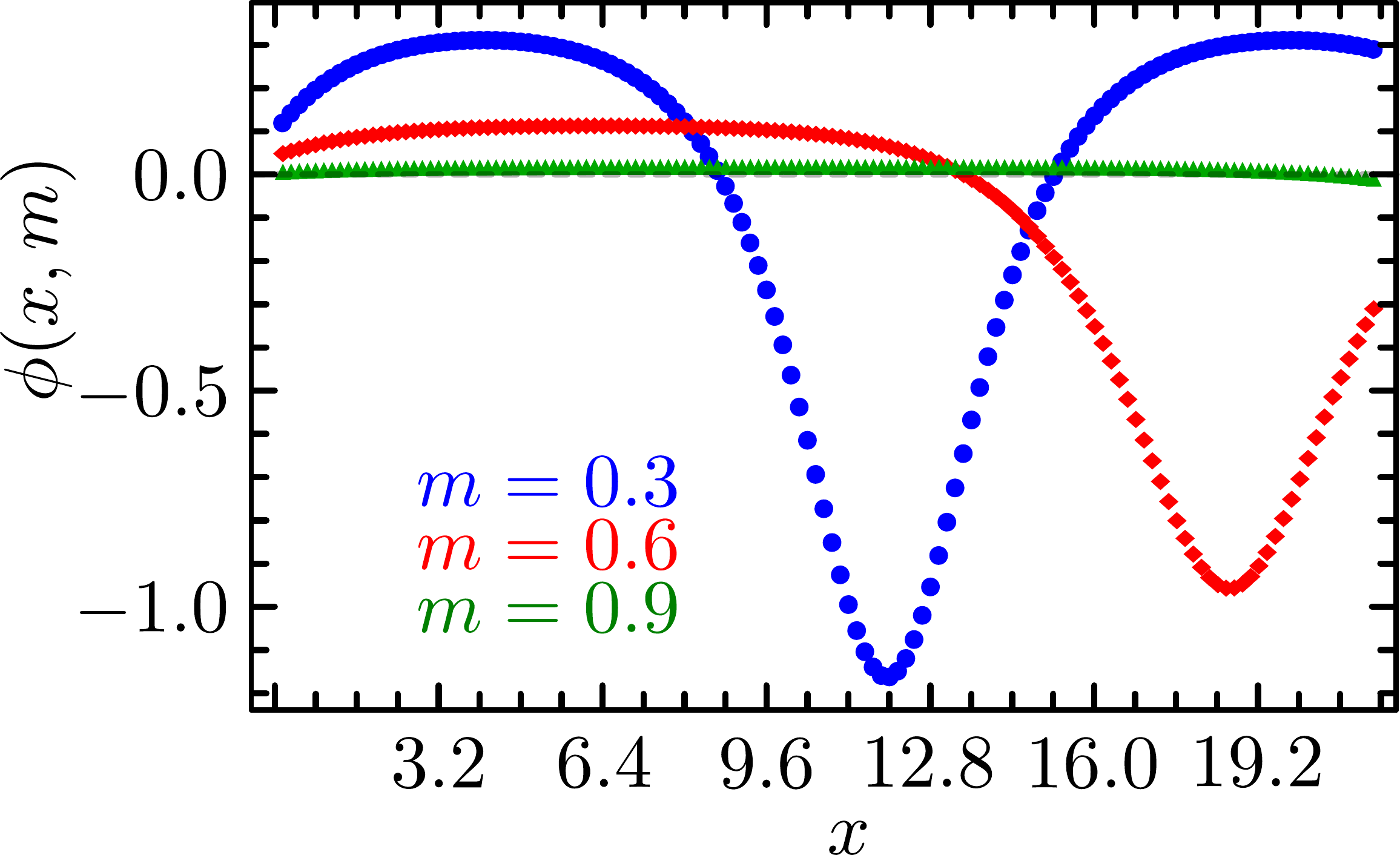} 
\caption{Variations of nonlinear Klein-Gordon fields $\phi(x,m)$ as a function of the position following Eq.~\eqref{eq.29}, Eq.~\eqref{eq.30.1}, and Eq.~\eqref{eq.30.2}. Note that the parameters are assumed to be $a = - 0.5$, and $b = 0.8$.}\label{fig:Fig4}
\end{figure}

The structure of these novel periodic kink solutions for the first case ($D^2 = \tfrac{1}{\sqrt{m}}$) is shown in Fig.~\ref{fig:Fig4} for distinct choices of the modulus $m$, and the parameters $a = - 0.5$, and $b = 0.8$. \\

{\bf Solution VIII}

It is well known that the $\phi^4$ Eq.~\eqref{eq.1} admits both $A/\dn(x,m)$ and $B\sn(x,m)/\dn(x,m)$ as exact solutions~\cite{ks18}, but we now show that remarkably, even their superposition is an exact solution of Eq.~\eqref{eq.1}. In particular, it is easy to check that
\begin{equation}\label{eq.34}
\phi = \frac{A \sqrt{1-m}}{\dn(\beta x,m)} + \frac{B\sqrt{m(1-m)}\sn(\beta x,m)}{\dn(\beta x,m)},
\end{equation}
is an exact solution of Eq.~\eqref{eq.1} provided
\begin{equation}\label{eq.35}
0 < m < 1,
\qquad
B = \pm A,
\qquad
2b A^2 = -\beta^2,
\qquad
a = \frac{(1+m)\beta^2}{2}.
\end{equation}
Unlike all the periodic kink solutions discussed above, solution VIII exists only if $b < 0, a > 0$. Further, unlike all the solutions discussed above, this solution is not an eigenstate of parity, i.e., while the first term is even under $x \rightarrow -x$, the second term is odd under parity.

We now discuss three complex $\cal{PT}$-invariant solutions with $\cal{PT}$-eigenvalue +$1$, three complex $\cal{PT}$-invariant solutions with $\cal{PT}$-eigenvalue $-1$ and two complex (but not $\cal{PT}$-invariant) solutions of Eq.~\eqref{eq.1}. \\

{\bf Solution IX}

It is easy to check that 
\begin{equation}\label{eq.36}
\phi = \frac{\sqrt{m}[A\cn(\beta x,m) + iB\sn(\beta x,m)]}{D+\dn(\beta x,m)},
\qquad
A,B,D > 0,
\end{equation}
is an exact complex $\cal{PT}$-invariant periodic solution with $\cal{PT}$-eigenvalue $+1$ of the $\phi^4$ Eq.~\eqref{eq.1} provided
\begin{equation}\label{eq.37}
2 b A^2 = (1-D^2)\beta^2,
\qquad
2 b B^2 = (1-m-D^2)\beta^2,
\qquad
a = -(2-m)\frac{\beta^2}{2}.
\end{equation}
Thus such a solution exists if either
\begin{equation}\label{eq.38}
D^2 > 1,	\;		b < 0, \qquad {\rm{or}} \qquad 0 < D^2 < 1-m,	 \;	  b > 0.
\end{equation}
Note that for both these cases $a < 0$. The corresponding well known complex $\cal{PT}$-invariant pulse solution of the $\phi^4$ Eq.~\eqref{eq.1} with $\cal{PT}$-eigenvalue $+1$ is
\begin{equation}\label{eq.40}
\phi = \sqrt{m}[A\cn(\beta x,m) + i\sn(\beta x,m)],
\end{equation}
provided
\begin{equation}\label{eq.41}
2 b A^2 = -\beta^2,
\qquad
a = -(2-m)\frac{\beta^2}{2}.
\end{equation}
Thus while the solution Eq.~\eqref{eq.40} only exists if $b < 0, a <0$, the new solution IX not only exists if $b < 0$ but also when $b > 0$ depending on the value of $D$. Thus the new solution IX covers far more parameter space in the $(a,b)$ plane compared to the well-known solution Eq.~\eqref{eq.40}. \\

{\bf Solution X}

It is easy to check that 
\begin{equation}\label{eq.42}
\phi = \frac{[A\dn(\beta x,m) + iB\sqrt{m}\sn(\beta x,m)]}{D+\cn(\beta x,m)},
\qquad
A, B > 0, D > 1,
\end{equation}
is an exact complex $\cal{PT}$-invariant periodic solution with $\cal{PT}$-eigenvalue $+1$ of the $\phi^4$ Eq.~\eqref{eq.1} provided
\begin{equation}\label{eq.43}
2 b A^2 = -(D^2-1)\beta^2,
\qquad
2 m b B^2 = -(m D^2+1-m)\beta^2,
\qquad
a = -(2m-1)\frac{\beta^2}{2}.
\end{equation}
Thus this solution exists only if $b < 0$. On the other hand $a > (<)$ 0 depending on if $m < 1/2$  $(> 1/2)$. In contrast, the corresponding well known complex $\cal{PT}$-invariant pulse solution of the $\phi^4$ Eq.~\eqref{eq.1} with $\cal{PT}$-eigenvalue $+1$ is
\begin{equation}\label{eq.44}
\phi = [A\dn(\beta x,m) + iB\sqrt{m}\sn(\beta x,m)],
\end{equation}
provided
\begin{equation}\label{eq.45}
B = \pm A,
\qquad
2 b A^2 = -\beta^2,
\qquad
a = -(2m-1)\frac{\beta^2}{2}.
\end{equation}
Thus for this solution $b < 0$ while $a > (< )$ 0 depending on if $m < (>)$ 1/2. However, the new solution X is very different from the well-known solution Eq.~\eqref{eq.44}. In particular, $B^2 = A^2$ for the old solution Eq.~\eqref{eq.44}, $B^2 \ne A^2$ for the new solution X. 

In the limit $m = 1$, both the solutions IX and X go over to the complex $\cal{PT}$-invariant hyperbolic solution with $\cal{PT}$-eigenvalue $+1$, i.e.
\begin{equation}\label{eq.46}
\phi = \frac{[A\sech(\beta x)+iB\tanh(\beta x)]}{D+\sech(\beta x)}\,,
~~A, B, D > 0\,,
\end{equation}
provided
\begin{equation}\label{eq.47}
2 b A^2 = -(D^2-1)\beta^2\,,~~2 b B^2 = -D^2\beta^2\,,~~
a = -\frac{\beta^2}{2}\,.
\end{equation}
Thus this solution only exists if $a< 0, b < 0$ and $D^2 > 1$. 

In contrast the corresponding well known complex $\cal{PT}$-invariant hyperbolic pulse solution of the $\phi^4$ Eq.~\eqref{eq.1} with $\cal{PT}$-eigenvalue $+1$ as
\begin{equation}\label{eq.48}
\phi = [A\sech(\beta x)+iB\tanh(\beta x)],
\end{equation}
provided
\begin{equation}\label{eq.49}
B = \pm A,
\qquad
2 b A^2 = -\beta^2,
\qquad
a = -\frac{\beta^2}{2}.
\end{equation}
Thus for this solution $b < 0, a <0$. We suspect that the hyperbolic solution Eq.~\eqref{eq.46} may not be new but could be related to the well known solution Eq.~\eqref{eq.48} by a translation. \\

{\bf Solution XI}

It is easy to check that 
\begin{equation}\label{eq.50}
\phi = \frac{[A+iB \sin(\beta x)]}{D+\cos(\beta x)},
\qquad
A, B > 0, D > 1, 
\end{equation}
is an exact complex $\cal{PT}$-invariant periodic solution with $\cal{PT}$-eigenvalue $+1$ of the $\phi^4$ Eq.~\eqref{eq.1} provided
\begin{equation}
2 b A^2 = -(D^2-1)\beta^2,
\qquad
2 b B^2 = -\beta^2,
a = \frac{\beta^2}{2}.
\end{equation}
Thus this solution exists only if $a > 0, b < 0$. 

We now discuss three complex $\cal{PT}$-invariant kinklike solutions with $\cal{PT}$-eigenvalue $-1$. \\

{\bf Solution XII}

It is easy to check that 
\begin{equation}\label{eq.51}
\phi = \frac{\sqrt{m}[A\sn(\beta x,m) + iB\cn(\beta x,m)]}{D+\dn(\beta x,m)},
\qquad
A, B, D > 0,
\end{equation}
is an exact complex $\cal{PT}$-invariant periodic solution with $\cal{PT}$-eigenvalue $-1$ of the $\phi^4$ Eq.~\eqref{eq.1} provided
\begin{equation}\label{eq.52}
2 b A^2 = (D^2-1+m)\beta^2,
\qquad
2 b B^2 = (D^2-1)\beta^2,
\qquad
a = -(2-m)\frac{\beta^2}{2}.
\end{equation}
It then follows that such a solution exists if either
\begin{equation}\label{eq.53}
D^2 > 1,	\; 	b > 0, \qquad {\rm{or}} \qquad 0 < D^2 < 1-m,	\; b < 0.
\end{equation}
Note that for both these cases $a < 0$. The corresponding well known complex $\cal{PT}$-invariant kink-like solution of the $\phi^4$ Eq.~\eqref{eq.1} with $\cal{PT}$-eigenvalue $-1$ is 
\begin{equation}\label{eq.55}
\phi = \sqrt{m}[A\sn(\beta x,m)+iB\cn(\beta x,m)],
\end{equation}
provided
\begin{equation}\label{eq.56}
B = \pm A,
\qquad
2 b A^2 = \beta^2,
\qquad
a = -(2-m)\frac{\beta^2}{2}.
\end{equation}
Thus while the well known solution Eq.~\eqref{eq.55} exists only if $b > 0, a <0$, the new solution XII exists for both $b > 0$ and $b < 0$. Thus the new solution covers far more parameter space in the $(a,b)$ plane and is very different from the well known solution Eq.~\eqref{eq.56}. \\

{\bf Solution XIII}

It is easy to check that 
\begin{equation}\label{eq.57}
\phi = \frac{[A\sqrt{m}\sn(\beta x,m) + iB\dn(\beta x,m)]}{D+\cn(\beta x,m)}, \qquad A, B > 0, D > 1,
\end{equation}
is an exact complex $\cal{PT}$-invariant periodic solution with $\cal{PT}$-eigenvalue $-1$ of $\phi^4$ Eq.~\eqref{eq.1} provided
\begin{equation}\label{eq.58}
2 m b A^2 = (1-m+m D^2)\beta^2, \qquad b B^2 = (D^2-1)\beta^2, \qquad a = -(2m-1)\frac{\beta^2}{2}.
\end{equation}
Thus this solution exists only if $b > 0$. On the other hand $a > (<)$ 0 depending on if $m < 1/2$ $(> 1/2)$. The corresponding well known complex $\cal{PT}$-invariant pulse solution of the $\phi^4$ Eq.~\eqref{eq.1} with $\cal{PT}$-eigenvalue $-1$ is 
\begin{equation}\label{eq.59}
\phi = [A\sqrt{m} \sn(\beta x,m)+iB\dn(\beta x,m)]\,
\end{equation}
provided
\begin{equation}\label{eq.60}
B = \pm A\,,~~2 b A^2 = \beta^2,
\qquad
a = -(2m-1)\frac{\beta^2}{2}.
\end{equation}

Thus for this solution $b > 0$ while $a > (< )$ 0 depending on if $m < (>)$ $1/2$. Note that the two solutions XII and Eq.~\eqref{eq.60} have very different behavior within a period.

In the limit $m = 1$, the solutions XII and XIII go over to the complex $\cal{PT}$-invariant hyperbolic solution with $\cal{PT}$-eigenvalue $-1$, i.e.
\begin{equation}\label{eq.61}
\phi = \frac{[A\tanh(\beta x) + iB\sech(\beta x)]}{D+\sech(\beta x)},
~~A,B,D > 0\,,
\end{equation}
provided
\begin{equation}\label{eq.62}
2 b A^2 = D^2 \beta^2,
\qquad
2 b B^2 = (D^2-1) \beta^2,
\qquad
a = -\frac{\beta^2}{2}.
\end{equation}
Thus this solution only exists if $a< 0, b > 0$ and $D^2 > 1$. 

In contrast, the corresponding well known complex $\cal{PT}$-invariant pulse solution of the $\phi^4$ Eq.~\eqref{eq.1} with $\cal{PT}$-eigenvalue $-1$ is
\begin{equation}\label{eq.63}
\phi = [A\tanh(\beta x)+iB\sech(\beta x)],
\end{equation}
provided
\begin{equation}\label{eq.64}
B = \pm A,
\qquad
2 b A^2 = \beta^2,
\qquad
a = -\frac{\beta^2}{2}.
\end{equation}
Thus for this solution $b > 0, a <0$. 

We suspect that the complex $\cal{PT}$-invariant hyperbolic solution Eq.~\eqref{eq.62} is not really new but related to the solution Eq.~\eqref{eq.64} by a translation. \\

{\bf Solution XIV}

It is easy to check that 
\begin{equation}\label{eq.65}
\phi = \frac{[A\sin(\beta x) + iB]}{D+\cos(\beta x)}, \qquad A, B > 0, D > 1,
\end{equation}
is an exact complex $\cal{PT}$-invariant periodic solution with $\cal{PT}$-eigenvalue $-1$ of $\phi^4$ Eq.~\eqref{eq.1} provided 
\begin{equation}\label{eq.66}
2 b A^2 = \beta^2,
\qquad
2 b B^2 = (D^2-1)\beta^2, \qquad a = \frac{\beta^2}{2}.
\end{equation}
Thus this solution exists only if $a, b > 0$. \\

{\bf Solution XV}

A novel complex (but which is not $\cal{PT}$-invariant) solution of the $\phi^4$ Eq.~\eqref{eq.1} is
\begin{equation}\label{eq.67}
\phi = \frac{A\sqrt{1-m}}{\dn(\beta x,m)} +\frac{iB\sqrt{m}\cn(\beta x,m)}{\dn(\beta x,m)},
\end{equation}
provided
\begin{equation}\label{eq.68}
B = \pm A,
\qquad
0 < m < 1,
\qquad
2b A^2 = -\beta^2,
\qquad
a = -(2m-1)\frac{\beta^2}{2}.
\end{equation}

{\bf Solution XVI}

Another novel complex (but which is not $\cal{PT}$-invariant) solution of the $\phi^4$ Eq.~\eqref{eq.1} is
\begin{equation}\label{eq.69}
\phi = \frac{A\sqrt{m}\cn(\beta x,m)}{\dn(\beta x,m)} +\frac{iB\sqrt{1-m}}{\dn(\beta x,m)},
\end{equation}
provided
\begin{equation}\label{eq.70}
B = \pm A,
\qquad
0 < m < 1,
\qquad
2b A^2 = \beta^2,
\qquad
a = -(2m-1)\frac{\beta^2}{2}.
\end{equation}

\begin{figure}[t]
\centering
\includegraphics[width=0.6\linewidth]{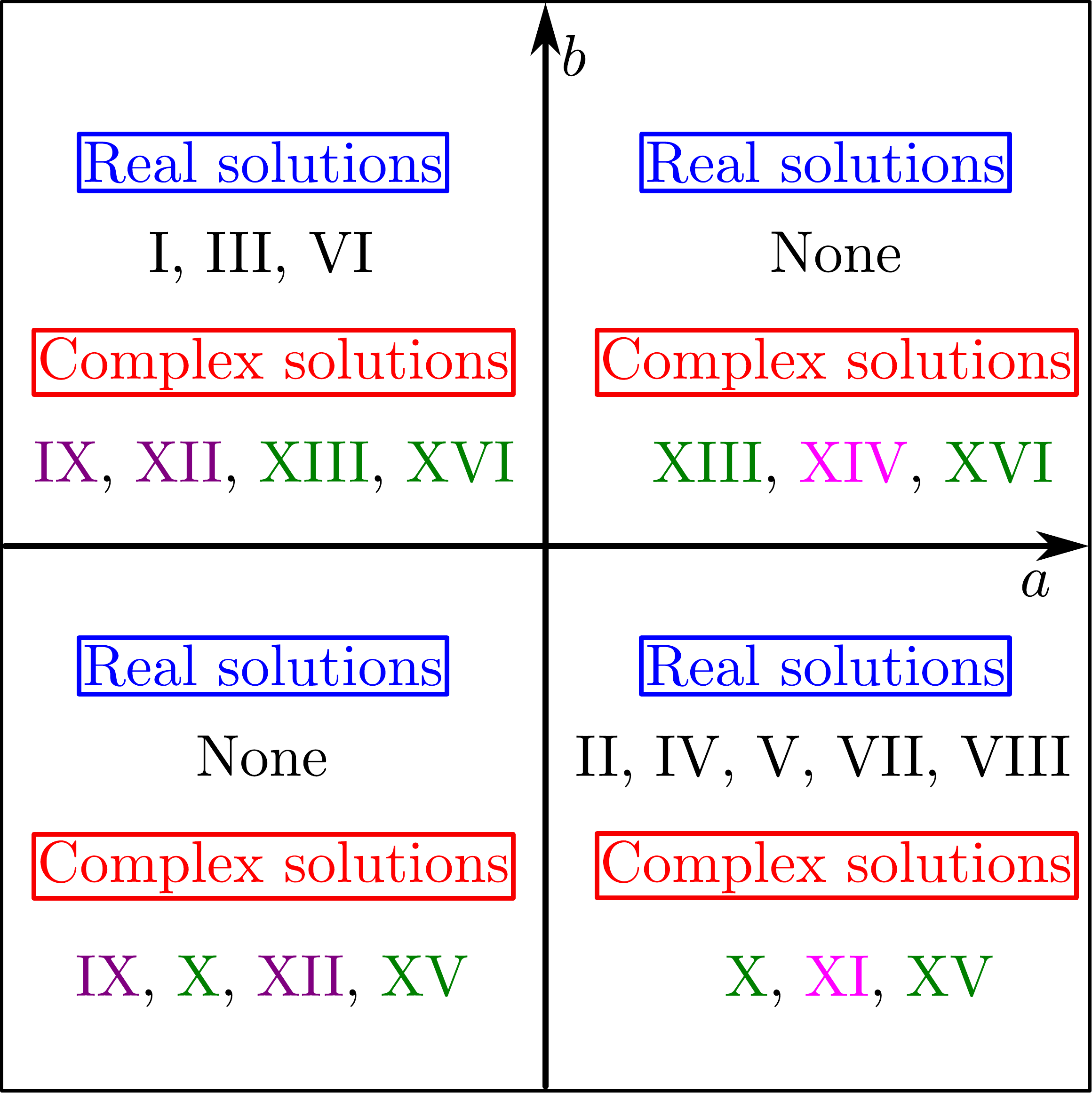} 
\caption{A graphical representation of the characteristic real and complex solutions (all sixteen) in the $ab$-plane.  Most of the complex solutions exist in two 
contiguous quadrants.}\label{fig:Fig5}
\end{figure}

\section{Conclusion \label{sec:sec_II}} 

In this paper, we have presented sixteen entirely new (spatially) periodic static solutions of the $\phi^4$ Eq.~\eqref{eq.1} and compared as well as contrasted them with the old known solutions. Remarkably, many of the new solutions occupy either bigger or different parameter space in the $(a,b)$ plane compared to the old solutions. In Fig. 5 we categorize both real and complex solutions in the $(a,b)$ plane. Real solutions exist in second and fourth quadrants only. Complex solutions exist in all four quadrants; however, each solution is shared in two quadrants (color coded), e.g., solutions IX and XII exist in both second and third quadrants.  Similarly, solutions XIII and XVI exist in both first and second quadrants.  The complex trigonometric solutions XI and XIV do not come as a pair but they exist alone in fourth and first quadrant, respectively. As far as the complex $\cal{PT}$-invariant solutions are concerned, we observe that the solution with $\cal{PT}$-eigenvalue +$1$ and the corresponding solution with $\cal{PT}$-eigenvalue $-1$ are related via $A \rightarrow iB$. In particular, the solutions IX and XII, X and XIII and XI and XIV  are related via $A \rightarrow iB$. However, the symmetry if any between the two complex but non-$\cal{PT}$ invariant solutions 
XV and XVI is not obvious. 

It would be fascinating to find the relevance of some of these solutions to physical problems. Since many other nonlinear equations have a deeper connection with the symmetric $\phi^4$ equation, we expect similar new solutions to exist in other nonlinear equations, including nonlocal nonlinear equations. We hope to report on such solutions soon.

\section{Acknowledgment} 

One of us (AK) is grateful to Indian National Science Academy (INSA) for the award of the INSA Honorary Scientist position at Savitribai Phule Pune University. The work at Los Alamos National Laboratory was carried out under the auspices of the US DOE and NNSA under contract No.~DEAC52-06NA25396.

\newpage 



\end{document}